\documentclass[11pt,superscriptaddress,aps,prd,preprint]{revtex4}
\usepackage[active]{srcltx}
\usepackage{graphicx}
\usepackage[utf8]{inputenc}
\usepackage{amsmath}
\usepackage{xcolor}
\usepackage{times,txfonts}
\usepackage{float}

\begin{document}

\title{Generalized Ellis-Bronnikov wormholes in asymptotically safe gravity}

\author{M. Nilton}
\affiliation{Universidade Federal do Cear\'{a}, Fortaleza, Cear\'{a}, Brazil}
\email{matheus.nilton@fisica.ufc.br}

\author{J. Furtado}
\affiliation{Universidade Federal do Cariri(UFCA), Av. Tenente Raimundo Rocha, \\ Cidade Universit\'{a}ria, Juazeiro do Norte, Cear\'{a}, CEP 63048-080, Brasil}
\email{job.furtado@ufca.edu.br}

\author{G. Alencar}
\affiliation{Universidade Federal do Cear\'{a}, Fortaleza, Cear\'{a}, Brazil}
\email{geova@fisica.ufc.br}

\author{R. R. Landim}
\affiliation{Universidade Federal do Cear\'{a}, Fortaleza, Cear\'{a}, Brazil}
\email{renan@fisica.ufc.br}

\date{\today}

\begin{abstract}
In this paper we study a class of wormhole solutions called generalized Ellis-Bronnikov wormholes in the context of asymptotically safe gravity (ASG). These solutions are characterized by two parameters: an even number $n$ and the wormhole throat radius $r_t$. The particular case $n=2$ recovers the usual Ellis-Bronnikov spacetime, which has already been addressed in the literature. We analyzed the nature of matter in the wormhole's throat, and in nearby regions, of these generalized solutions with $n>2$, using three curvature scalars in the ASG approach, namely, the Ricci scalar, squared Ricci and the Kretschmann scalar. We have shown that the ASG leads to corrections in the matter at the wormhole's throat only for the $n=4$ case. For the squared Ricci and the Kretschmann we find that exotic matter is always necessary, as previously found for the usual Ellis-Bronnikov. However, for the Ricci scalar case, we found that ordinary matter is allowed at the throat. Therefore, the generalized Ellis-Bronnikov wormhole provides to the possibility of having ordinary matter at the throat in the context of asymptotically safe gravity.
\end{abstract}

\maketitle

\section{Introduction}

\hspace{0.4cm} When trying to describe a particle model using general relativity, Einstein and Rosen arrived at the concept of a bridge, which can be interpreted as a tunnel connecting different universes, or asymptotically flat regions of the same universe \cite{Einstein:1935tc}. This type of solution became known in the literature as Einstein-Rosen bridges and is just one example of a class of solutions that was called wormholes by Misner and Wheeler \cite{Misner:1957mt}. These solutions were speculated to be an alternative to fast travel between distant points of spacetime. However, the study of a general construction of traversable wormholes was made only by Morris and Thorne \cite{Morris:1988cz}, where they imposed a series of physical constraints on an ansatz metric, and then they determined what conditions led to specifically humans being able to cross it. Morris and Thorne showed that the existence of a traversable wormhole necessarily needs the presence of matter, such that, in the throat of the wormhole, the tension must exceed the total energy density. This kind of matter was called ``exotic'', and they satisfy the property of the positivity of the exoticity function $\xi_{e}=-(\rho+p_r)/\rho$ \cite{Morris:1988cz}. Putting in terms of the state parameter $\omega$, defined by $p_r=\omega\rho$, the exotic condition implies that $\omega<-1$, and this is what characterizes phantom matter. The presence of exotic matter has been seen in a wide variety of wormholes \cite{Sushkov:2005kj, Lobo:2005us, Garattini:2019ivd, Jusufi:2020rpw, Alencar:2021ejd, Oliveira:2021ypz, Carvalho:2021ajy} and what we can do to try to eliminate this problem, is appeal to modified theories in a tentative to construct wormholes without the exigence of exotic matter \cite{Richarte:2007zz, Matulich:2011ct, Richarte:2009zz, MontelongoGarcia:2011ag, Ovgun:2018xys, Chew:2016epf, Chew:2018vjp}.

We can expect that quantum effects in the gravitational field provide the existence of wormholes with ordinary matter satisfying the energy conditions. The problem then would be how we describe quantum corrections for gravitational systems. When we try to quantize gravitation along the lines of quantum field theory, this leads to the appearance of divergences that can not be eliminated using the usual renormalization techniques. However, the asymptotically safe conjecture proposed by Weinberg \cite{Weinberg} proved to be quite promising as it provides a renormalizability condition that saves the theory from divergences and makes it predictive. According to this conjecture, the theory is free from ultraviolet (UV) divergences if it admits non-Gaussian fixed points such that all the running coupling constants of the theory tend to it in the UV limit. A theory that satisfies this condition will be said asymptotically safe.

The asymptotically safe formalism has been successfully applied to gravitational systems \cite{Saueressig:2015xua,Pawlowski:2018swz,Platania:2019kyx,Bosma:2019aiu,Ishibashi:2021kmf} solving the singularity problem in black holes spacetimes and relating to other methods to describe quantum gravity \cite{Borissova:2020knn,Chojnacki:2021xtr,Giacchini:2021pmr,Chojnacki:2021ves}. The existence of a fixed point for gravity theory has been found in several different considerations \cite{Reuter:1996cp, Lauscher:2001ya, Litim:2003vp, Machado:2007ea, Benedetti:2009rx, Manrique:2011jc, Christiansen:2012rx, Morris:2015oca, Demmel:2015oqa, Platania:2017djo, Christiansen:2017bsy, Falls:2018ylp, Narain:2009fy, Oda:2015sma, Eichhorn:2017ylw, Eichhorn:2019yzm, Reichert:2019car, Daas:2020dyo}. This method consists in determining the gravitational flow $\Gamma_k[g_{\mu\nu}]$, which is a solution of the exact renormalization group equation (ERGE) \cite{Reuter:1996cp}
\begin{equation}
k\partial_k\Gamma_k=\frac{1}{2}\textrm{Tr}[(\Gamma^{(2)}_k+\mathcal{R}_k)^{-1}k\partial_k\mathcal{R}_k],
\end{equation}
where $k$ is the renormalization group scaling parameter, $\Gamma^{(2)}_k$ is the Hessian of the $\Gamma_k$ and the cutoff function $\mathcal{R}_k$ is introduced in order to eliminate the infrared (IR) divergences, suppressing the small moment modes, i.e, those with $p^2<<k^2$, and has the requirement to be a quadratic function of the moment \cite{Reuter:1996cp}. To reduce the difficulty of solving ERGE, we usually appeal to the truncation methods, which consist of projecting the basis of the theory of space, where the flow is defined, in a subspace that accounts all the essential physics. In the gravitational case, the Einstein-Hilbert truncation consists in writing the flow in the basis $\sqrt{g}$, $\sqrt{g}R$, which gives the following form for the antiscreening running coupling constant when we disregard the cosmological influence \cite{Reuter:1996cp,Bonanno:2000ep}
\begin{equation}
\label{running-coupling-constant}
G(k)=\frac{G(k_0)}{1+\omega G(k_0)(k^2-k_0^2)},
\end{equation}
where $\omega=\frac{4}{\pi}(1-\pi^2/144)$ and the moment of reference $k_0$ can be, without loss of generality, considered zero since $G(k_0\rightarrow0)\rightarrow G_0$, where $G_0$ is the low energy measure of Newton's constant \cite{Reuter:1996cp,Bonanno:2000ep}. Therefore, the consequences of this quantization approach can be done by improving the usual Newton's constant $G_0$ into a function of the renormalization group parameter $G(k)$ given by Eq.\eqref{running-coupling-constant}. This improvement can be placed directly in the classical solutions \cite{Reuter:2003ca} or in the classical field equation \cite{Reuter:2003ca,Reuter:2010xb,Falls:2010he}. However, it can be shown that the improvement of the running coupling constant can be done in a more physical way by improving the $G_0\rightarrow G(k)$ in the classical action \cite{Moti:2018rho}.

Furthermore, it is very convenient to identify the cutoff moment k with some quantity of physical interest in the theory. As was argued in \cite{Moti:2018rho} the most suitable choice for gravitational systems is to make the following identification $k=\xi/\chi$, where $\xi$ is a dimensionless constant and $\chi$ is a function of the invariants of curvature constructed by the components of the Riemann tensor, which describe the tidal forces. This leads to the following form for the gravitational running coupling constant
\begin{equation}\label{Gchi}
    G(\chi)=\frac{G_0}{1+f(\chi)},
\end{equation}
where $G_0$ is the Newton's constant and $f(\chi)=\xi/\chi$ is a function of the curvature invariants. The problem with this identification consists of a choice of the parameter $\chi$, which can be done by making some physical considerations such as geodesic congruence \cite{Moti:2019mws}. However, for simple models, it is common to consider $\chi=R$, $\chi=(R_{\alpha\beta}R^{\alpha\beta})^{1/2}$, and $\chi=(R_{\alpha\beta\gamma\delta}R^{\alpha\beta\gamma\delta})^{1/2}$ as studied for the FRW cosmological solution and other non-vacuum solutions \cite{Moti:2018rho, Moti:2019mws, Babic:2004ev, Domazet:2010bk, Domazet:2012tw}. With this, we can obtain the improved action by putting the gravitational running coupling constant \eqref{Gchi} in place of Newton's constant. This leads to modified field equations of the form \cite{Moti:2020whf} 
\begin{equation}\label{ASG2}
    G_{\mu\nu}=8\pi G(\chi) T_{\mu\nu}+G(\chi)X_{\mu\nu},
\end{equation}
where $X_{\mu\nu}$ is a covariant tensor related to the derivation of $G(\chi)$ with respect to the metric, which is given by
\begin{eqnarray}
\nonumber X_{\mu\nu}&=&\left(\nabla_{\mu}\nabla_{\nu}-g_{\mu\nu}\Box\right)G(\chi)^{-1}-\frac{1}{2}\left[R\mathcal{K}(\chi)\frac{\delta\chi}{\delta g^{\mu\nu}}+\partial_k\left(R\mathcal{K}(\chi)\frac{\partial\chi}{\partial(\partial_k g^{\mu\nu})}\right)\right.\\
&&\left.+\partial_k\partial_{\lambda}\left(R\mathcal{K}(\chi)\frac{\partial\chi}{\partial(\partial_{\lambda}\partial_k g^{\mu\nu})}\right)\right],
\end{eqnarray}
with $\mathcal{K}(\chi)=\frac{2}{G(\chi)^2}\frac{\partial G(\chi)}{\partial\chi}$. The tensor $X_{\mu\nu}$ describes the dynamics of quantum corrections encoded in $G(\chi)$, and can be interpreted as an effective energy-momentum tensor related to the 4-momentum of running coupling constant $G(\chi)$. Note that this brings naturally a quantum nature to the matter, and therefore, the Asymptotically Safe Gravity (ASG) treats the matter described by the energy-momentum tensor $T_{\mu\nu}$ as the quantum one. This feature turns more clear if we rewrite the modified field equation \eqref{ASG2} in the form
\begin{equation}
    \frac{1}{G(\chi)}\left(G_{\mu\nu}-T^{eff}_{\mu\nu}\right)=8\pi T_{\mu\nu}.
\end{equation}

In particular, the ASG approach has been applied in the context of wormholes in order to investigate if these quantum effects provide traversable solutions with ordinary matter satisfying the energy conditions. In the Ref.\cite{Moti:2020whf}, the authors studied the effects of ASG on the conditions of traversability in two classes of wormholes, the spherical and the pseudospherical, considering an anisotropic fluid as the source that obeys a linear state equation with the state parameter being constant. They found that, in this case, only the pseudospherical solutions can be traversable with ordinary matter, satisfying the null energy condition in the throat. More recently, the authors in  \cite{Nilton:2022cho} investigated the traversability conditions and the presence of exotic matter along the same lines as Ref.\cite{Moti:2020whf}, but considering a more general case where the state parameter is dependent on the position. They found that this general case gives the possibility of traversability with ordinary matter in the spherical case too. Also, the ASG approach has been applied in specific wormhole spacetimes such as the Ellis-Bronnikov \cite{Alencar:2021enh} where the authors found that the throat can never be sourced by ordinary matter. Also, the Schwarzschild and Schwarzschild-like wormholes were studied in the context of ASG \cite{Nilton:2021pyi}, where the authors found the possibility of ordinary matter as the source for the wormhole at the throat as well as in other regions of space.

Some time ago a class of wormhole solutions based in the Ellis-Bronnikov spacetime \cite{Ellis:1973yv,Bronnikov:1973fh}  was proposed in \cite{Kar:1995jz}, as an attempt to circumvent the problem of exotic matter. These solutions were called generalized Ellis-Bronnikov wormholes which have been studied in the context of general relativity \cite{Kar:1995jz}. The authors investigated the resonances of the propagation of scalar waves in this family of wormholes. Furthermore, they show that these classes of wormholes only can be supported by exotic matter, because we always have the violations of the null energy condition and the weak energy condition in the classical context. More recently, these solutions were revisited in Ref.\cite{DuttaRoy:2019hij} where the authors showed that the necessity of extra matter beyond phantom to support the geometries of these generalized solutions. In particular, the quasinormal modes were studied. Also, this metric was recently studied in the context of braneworlds, where the authors in \cite{Sharma:2021kqb} embedded these solutions in a five-dimensional warp braneworld and this brings the possibility that the weak energy condition is satisfied. The stability of these generalized solutions considering axial gravitational perturbations was made in \cite{Roy:2021jjg}. In this work, we verify if it is possible to have a throat without phantom in the generalized Ellis-Bronnikov wormholes in the context of Asymptotically Safe Gravity (ASG). In particular, we check the possibility of having only ordinary matter in the throat of these generalized solutions. Furthermore, we analyze the validity of the radial energy conditions at the throat. As we will see, the necessity of ordinary matter is found for one of these generalized solutions when we use the Ricci scalar model in the ASG approach.

This paper is organized as follows: In the next section we discuss the generalized Ellis-Bronnikov solutions in the context of ASG. Also we study which matter is necessary in the throat of these generalized solutions. Then, we verify if the radial energy conditions at the throat is satisfied. We will use the Ricci scalar, squared Ricci scalar and the Kretschmann scalar, respectively, to define the improvement done by the cutoff function. In section III we give the final remarks and the conclusions.

\section{Generalized Ellis-Bronnikov wormhole solution in ASG}
Let us consider the spherically Morris-Thorne wormhole, whose metric is given by \cite{Morris:1988cz}
\begin{equation}
   ds^2 = e^{2\Phi(r)} dt^2 - \frac{dr^2}{1-b(r)/r} -r^2d\Omega_{2(s)} \ , \label{SSM}
\end{equation}
where $e^{2\Phi(r)}$ is the redshift function, $b(r)$ the shape function and $d\Omega_{2(s)} = d\theta^{2} + \sin^{2}{\theta}d\phi^{2}$ is the line element of a 2-sphere.

Considering an anisotropic fluid with $T_{\nu}^{\mu}=\textrm{Diag}[\rho(r),-p_r(r),-p_l(r),-p_l(r)]$ that generates the wormhole geometry we obtain the modified field equations \cite{Moti:2020whf}
\begin{eqnarray}
     8\pi G_0 \rho &=&\bigl(1+f\bigr) \frac{b^{'}}{r^2} -\left(1-\frac{b}{r}\right) \left(f''+\frac{2}{r} f'\right)+ \frac{b^{'}r-b}{2r^2}f'  \label{IEQ-tt-s}\\
      8\pi G_0 p_r  &=& -\bigl(1+f\bigr) \left(\frac{b}{r^3}-\frac{2\Phi^{'}}{r}\left(1-\frac{b}{r}\right) \right)+ \left(1-\frac{b}{r}\right) \left( \Phi^{'}+\frac{2}{r}\right) f'   \label{IEQ-rr-s} \\
    \nonumber8\pi G_0 p_l  &=& -\bigl(1+f\bigr) \left( \frac{b'r-b}{2r^2} \left(\Phi'+\frac{1}{r} \right)-\left(1-\frac{b}{r}\right)\left(\Phi''+\Phi'^2+\frac{\Phi'}{r}\right) \right)\\
     &&+ \left(1-\frac{b}{r}\right)\left( \left( \Phi^{'}+\frac{1}{r}\right)f'+f''\right)-\frac{b'r-b}{2r^2}f' \ , \label{IEQ-pp-s}
  \end{eqnarray}

where the prime means derivative with respect to the radial coordinate $r$. It is important to note that ASG provides quantum corrections to the matter sector as well, encoded in the constant $\xi$. If we look these equations for $\rho$, $p_r$ and $p_l$, we can separate them in a part independent of the ASG (without corrections) and in other part dependent of the ASG (the terms that are multiplied by $f$, and consequently, the constant $\xi$). Indeed, if we take the limit $\xi\rightarrow 0$ in the Eqs.(8), (9) and (10), these quantities tends to their classical expressions.  In this way, we can think that the extra matter necessary to support the geometries of the generalized Ellis-Bronnikov wormholes comes from the quantum corrections in the matter sector, which in this case is introduced through the ASG. We will check if these corrections in the matter provides a throat without phantom. In particular, the possibility of having only ordinary matter in the throat of generalized Ellis-Bronnikov wormholes and in regions nearby of the throat. The generalized Ellis-Bronnikov wormhole \cite{Kar:1995jz,DuttaRoy:2019hij} is defined by the following conditions on the shape and redshift functions
\begin{eqnarray}
\label{shape-function}b(r)&=&r-r^{3-2 n} \left(r^n-r_t^n\right)^{2-\frac{2}{n}},\\
\Phi(r)&=&0 ,
\end{eqnarray}
with $r_t$ being the wormhole throat radius. The condition $\Phi(r)=0$ imply a zero--tidal wormhole, and the parameter $n$ is allowed to assume only even values \cite{Kar:1995jz,DuttaRoy:2019hij} with $n=2$ being the usual Ellis-Bronnikov wormhole spacetime. A peculiar feature of this family of wormholes is that the throat must be sourced by matter with $\omega=-1$, with the exception of the usual Ellis-Bronnikov spacetime, where we have only phantom as source \cite{DuttaRoy:2019hij}. This is due to the fact that these generalized Ellis-Bronnikov wormholes has the necessity of extra matter, which is null in the $n=2$ case. Now, we will verify what kinds of matter is necessary to support these family of wormholes in the ASG approach.

Under these considerations the field equations simplify to
\begin{eqnarray}
\label{rho1}\nonumber\kappa\rho&=&r^{-2 (n+1)} \left(r^n-r_t^n\right)^{-2/n} \left\{-r^2 r_t^{2 n} \left[r^2 f''-(n-3) r f'+(3-2 n) f-2 n+3\right]\right.+\\
\nonumber&&\left.+r^{n+2} r_t^n \left[2 r^2 f''-(n-5) r f'-2 (n-2) f-2 n+4\right]+r^{2 n} \left[(f+1) \left(r^n-r_t^n\right)^{2/n}-r^2 \left(r^2 f''+2 r f'+f+1\right)\right]\right\},\\\\
\label{pr1}\kappa p_r&=&2 r^{1-2 n} f' \left(r^n-r_t^n\right)^{2-\frac{2}{n}}+\frac{(-f-1) \left(r-r^{3-2 n} \left(r^n-r_t^n\right)^{2-\frac{2}{n}}\right)}{r^3},\\
\kappa p_l&=&r^{-2 n} \left(r^n-r_t^n\right)^{\frac{n-2}{n}} \left(r^{n+1} \left(r f''+f'\right)+r_t^n \left(r \left((n-2) f'-r f''\right)+(n-1) f+n-1\right)\right),
\end{eqnarray}
with $\kappa=8\pi G_0$. Now we must set the function of the invariants of curvature $f$. Therefore, as pointed out previously, three choices are common for the curvature invariants, namely, the Ricci scalar, squared Ricci and Kretschmann scalar. Hence we will define $f_1=\xi R$, $f_2=\xi (R^{\mu\nu}R_{\mu\nu})^{1/2}$ and $f_3=\xi(R_{\mu\nu\lambda\rho}R^{\mu\nu\lambda\rho})^{1/2}$, so that
\begin{eqnarray}
\label{f1}f_1&=&-\frac{2 \xi}{r^2}  \left[-\left(2-\frac{2}{n}\right) n r^{2-n} \left(r^n-r_t^n\right)^{1-\frac{2}{n}}-(3-2 n) r^{2-2 n} \left(r^n-r_t^n\right)^{2-\frac{2}{n}}+1\right],\\
\label{f2}\nonumber f_2&=&\left\{2 r^{-4 (n+1)} \left(r^n-r_t^n\right)^{-4/n} \left[-2 (3 (n-3) n+8) r^{n+4} r_t^{3 n}+(n (3 n-8)+6) r^4 r_t^{4 n}+\right.\right.\\
\nonumber&&+2 (n-3) r^{3 n+2} r_t^n \left(r^2-\left(r^n-r_t^n\right)^{2/n}\right)+r^{4 n} \left(r^2-\left(r^n-r_t^n\right)^{2/n}\right)^2+\\
&&\left.\left.+r^{2 n+2} r_t^{2 n} \left(2 (n-2) \left(r^n-r_t^n\right)^{2/n}+3 ((n-4) n+5) r^2\right)\right]\right\}^{1/2}\xi,\\
\label{f3}\nonumber f_3&=&\left\{r^{-4 (n+1)} \left(r^n-r_t^n\right)^{-4/n} \left[-4 ((n-2) n+2) r^{n+4} r_t^{3 n}+(2 (n-2) n+3) r^4 r_t^{4 n}+4 r^{3 n+2} r_t^n \left(\left(r^n-r_t^n\right)^{2/n}-r^2\right)\right.\right.+\\
&&\left.\left.+r^{4 n} \left(r^2-\left(r^n-r_t^n\right)^{2/n}\right)^2+2 r^{2 n+2} r_t^{2 n} \left(((n-2) n+4) r^2-\left(r^n-r_t^n\right)^{2/n}\right)\right]\right\}^{1/2}\xi.
\end{eqnarray}

Now, we use these three scalars of curvature as improvement in order to analyze what kind of matter is necessary in the throat and the validity of the radial energy conditions. To do this we must substitute the cutoff fuctions defined above into (\ref{rho1}) and (\ref{pr1}). For the sake of clarity we will address each one of the cutoff functions separately. 

\subsection{Ricci scalar}

For the function of the Ricci scalar, defined by (\ref{f1}) we have

\begin{eqnarray}
\nonumber\kappa \rho&=&r^{-4 (n+1)} \left(r^n-r_t^n\right)^{-4/n} \left\{-2 r^{3 n+2} r_t^n \left(2 \left(n \left(n^2+n-2\right)-6\right) \xi  r^2+\left((n-2) r^2-2 (n-5) \xi \right) \left(r^n-r_t^n\right)^{2/n}\right)\right.+\\
\nonumber&&-4 ((n-1) n (12 n-11)-6) \xi  r^{n+4} r_t^{3 n}+6 (2 n-3) (2 (n-1) n+1) \xi  r^4 r_t^{4 n}+\\
\nonumber&&+r^{4 n} \left(\left(r^n-r_t^n\right)^{2/n}-r^2\right) \left(\left(r^2-2 \xi \right) \left(r^n-r_t^n\right)^{2/n}+6 \xi  r^2\right)+\\
&&\left.+r^{2 n+2} r_t^{2 n} \left(4 (n (7 (n-1) n-3)-6) \xi  r^2+\left((2 n-3) r^2-4 (n-3) \xi \right) \left(r^n-r_t^n\right)^{2/n}\right)\right\},\label{general-energy-density}\\
\nonumber\kappa p_r&=&r^{-4 (n+1)} \left(r^n-r_t^n\right)^{-4/n} \left\{-4 (5 n (2 n-3)-1) \xi  r^{n+4} r_t^{3 n}+2 (2 n (4 n-7)+3) \xi  r^4 r_t^{4 n}+\right.\\
\nonumber&&-2 r^{3 n+2} r_t^n \left(2 (n (2 n-1)-7) \xi  r^2+\left(2 (n+1) \xi +r^2\right) \left(r^n-r_t^n\right)^{2/n}\right)+\\
\nonumber&&-r^{4 n} \left(\left(r^n-r_t^n\right)^{2/n}-r^2\right) \left(\left(r^2-2 \xi \right) \left(r^n-r_t^n\right)^{2/n}-6 \xi  r^2\right)+\\
&&\left.+r^{2 n+2} r_t^{2 n} \left(4 (n (8 n-9)-8) \xi  r^2+\left(4 n \xi +r^2\right) \left(r^n-r_t^n\right)^{2/n}\right)\right\} \label{general-radial-pressure}.
\end{eqnarray}
Note that for $n=2$ we recover the results presented in \cite{Alencar:2021enh} where the authors studied the usual Ellis-Bronnikov wormhole solutions in the context of ASG. Here, we focus only in the generalized solutions with $n\geq4$.

Now, we will show that, using the Ricci scalar improvement in the generalized Ellis-Bronnikov wormhole, we have corrections in the throat only for the $n=4$ case, which gives a different result for energy density $\rho$, while for $n\geq6$ cases we always have the same result. For the  radial pressure, we will have the same result for all allowed $n\geq4$. For this, we rewrite the Eqs. \eqref{general-energy-density} and \eqref{general-radial-pressure} in the convenient form
\begin{eqnarray}
\kappa\rho&=&\frac{1}{r^{2}}-\frac{2\xi}{r^{4}}+r^{-2(1+n)}(r^{n}-r_{t}^{n})^{\frac{-2+n}{n}}(-r^{n}(r^{2}-8\xi)+3r_{0}^{n}(r^{2}-4\xi))-4n^{2}r^{-4n}r_{t}^{n}(r^{n}-r_{t}^{n})^{\frac{-4+n}{n}}(r^{2n}+8r^{n}r_{t}^{n}-15r_{t}^{2n})\xi \nonumber\\
&&-6 r^{-4n}(r^{n}-3r_{t}^{n})(r^{n}-r_{t}^{n})^{\frac{-4+n}{n}}(r^{2n}+r_{t}^{2n})\xi-4n^{3}r^{-4n}r_{t}^{n}(r^{n}-r_{t}^{n})^{\frac{-4+n}{n}}(r^{2n}-6r^{n}r_{t}^{n}+6r_{t}^{2n})\nonumber \\
&&-2n r^{-2(1+n)}r_{t}^{n}(r^{n}-r_{t}^{n})^{-2+n}{n}(r^{2}-2\xi)+4n r^{-4n}r_{t}^{n}(r^{n}-r_{t}^{n})^{\frac{-4+n}{n}}(2r^{2n}-r^{n}r_{t}^{n}-12r_{t}^{2n})\xi,
\label{energy-density}
\end{eqnarray}
and
\begin{eqnarray}
\kappa p_{r}&=&-\frac{1}{r^{2}}+\frac{2\xi}{r^{4}}-8n^{2}r^{-4n}r_{0}^{n}(r^{n}-2r_{t}^{n})(r^{n}-r_{t}^{n})^{2-4/n}\xi-2r^{-4n}(r^{n}-3r_{0}^{n})(r^{n} -r_{t}^{n})^{2-4/n}(3r^{n}+r_{t}^{n})\xi+ \nonumber\\
&&r^{-2(1+n)}(r^{n}-r_{t}^{n})^{\frac{-2+n}{n}}(r^{2+n}-r^{2}r_{t}^{n}+4r^{n}\xi)+4n r^{-4n}r_{t}^{n}(r^{n}-7r_{t}^{n})(r^{n}-r_{t}^{n})^{2-4/n}\xi-4n r^{-2(1 + n)}r_{t}^{n}(r^{n}-r_{t}^{n})^{\frac{-2+n}{n}}\xi.
\label{radial-pressure}
\end{eqnarray}
Firstly, looking at the Eq.\eqref{energy-density}, we can see that for $n\geq6$, all the terms of the type $r^{n}-r_{t}^{n}$ will have a positive exponent, and therefore, in the limit $r\rightarrow r_{t}$ these terms tend to zero leading to a vanishing of all the terms that depend on $n$, leaving only the first two parts of this equation, which gives us
\begin{equation}
\kappa\rho(r_{t})=\frac{r_{t}^{2}-2\xi}{r_{t}^{4}}\,\,\,\textrm{for}\,n\geq6.
\end{equation}
On the other hand, for $n=4$, all the terms of the type $(r^{n}-r_{t}^{n})$ with exponent $\frac{-4+n}{n}$, will give us the unity while those with exponent $\frac{-2+n}{n}$ will be a positive exponent, which vanishes in the limit $r\rightarrow r_{t}$. With this, we have
\begin{equation}
\kappa\rho(r_{t})=\frac{r_{t}^{2}-26\xi}{r_{t}^{4}}\,\,\,\textrm{for}\,n=4.
\end{equation}
For radial pressure, looking at the Eq. \eqref{radial-pressure}, it is easy to see that for $n\geq4$ all the terms of the type $(r^{n}-r_{t}^{n})$ will have a positive exponent and, in the limit $r\rightarrow r_{t}$, these terms are zero, vanishing all the terms that are dependent of $n$. Therefore, we have
\begin{equation}
\kappa p_{r}(r_{t})=-\frac{(r_{t}^{2}-2\xi)}{r_{t}^{4}}\,\,\,\textrm{for all}\,n\geq4.
\end{equation}

Note that, for generalized geometries with $n\geq6$, we have matter in the throat that satisfies the state equation $p_r(r_t)=-\rho(r_t)$. This is the same result predicted by \cite{Kar:1995jz,DuttaRoy:2019hij}. Then, we analyze the radial energy conditions and the necessary matter in the throat of generalized Ellis-Bronnikove wormhole only for the case $n=4$, that gives a different result reported in \cite{Kar:1995jz,DuttaRoy:2019hij}.

The radial energy conditions in the throat of the wormhole, so that for $r=r_t$ and $n=4$, gives:
\begin{eqnarray}
\kappa\rho&=&\frac{1}{r_t^4}\left(r_t^2-26 \xi\right),\\
\kappa p_r&=&-\frac{1}{r_t^4}\left(r_t^2-2 \xi\right),\\
\kappa(\rho+p_r)&=&-\frac{24 \xi }{r_t^4}.
\end{eqnarray}
The above equations shows us that, in the throat, the Null Energy Condition (NEC), i.e. $p_r+\rho>0$, is never satisfied. Therefore, NEC can not be satisfied in the throat, as the usual case \cite{DuttaRoy:2019hij}. Although we can have matter in the throat such that the energy density can be positive, as long as we have a throat radius satisfying $r_t>\sqrt{26\xi}$. Therefore, in order to satisfy Weak Energy Condition (WEC) and Dominant Energy condition (DEC) in the throat of the wormhole, which require $\rho\geq0$, we must have $r_t^2>26\xi$. However, the WEC can not be satisfied once we always have $p_r+\rho<0$. Also, DEC can not be satisfied, because in the range where the energy density is positive, $r_t>\sqrt{26\xi}$, we can easily see that we have $|p_r|>\rho$. Therefore, we conclude that the radial energy conditions at throat of generalized Ellis-Bronnikov wormholes can not be satisfied in the ASG context using the Ricci scalar as improvement.  

Now, let us study the presence of cosmological exotic matter in the generalized Ellis-Bronnikov wormhole with $n=4$ in the ASG scenario. To perform such investigation, we need to evaluate the state parameter $\omega(r)=p_r/\rho$. At the throat of the wormhole we have
\begin{eqnarray}
\omega(r_t)=\frac{2\xi-r_{t}^{2}}{r_t^2-26 \xi }.
\end{eqnarray}
Analyzing this as a function of $r_{t}$, we can easily obtain
\begin{equation}
\begin{cases}\label{closer_0}
\mbox{Phantom energy}\quad\quad\quad\quad\omega<-1 & \mbox{if}\quad r_{t}>\sqrt{26\xi},\\
\mbox{Matter with}\quad\quad\omega\geq1 & \mbox{if}\quad \sqrt{14\xi}\leq r_t<\sqrt{26\xi},\\
\mbox{Ordinary matter}\quad\quad 0<\omega<1 & \mbox{if}\quad \sqrt{2\xi}<r_t<\sqrt{14\xi},\\
\mbox{Matter with}\quad\quad -1/3<\omega<0 & \mbox{if}\quad r_t<\sqrt{2\xi}.
\end{cases}
\end{equation}

We can see that, contrary to what is predicted by the general relativity, we can have several kinds of matter, depending on the value of the throat radius $r_t$ that we choose, including cosmological matter with $-1/3<\omega<0$ and $\omega>1$, phantom ($\omega<-1$) and interestingly ordinary matter ($0<\omega<1$) too. Therefore, although the energy conditions are violated, the throat of a generalized Ellis-Bronnikov wormhole can be sourced only by ordinary matter for the case $n=4$, and, for this, the value of the throat radius it has to be between $\sqrt{2\xi}$ and $\sqrt{14\xi}$. 

\begin{figure}[ht!]
    \centering
     \includegraphics[width=0.6\textwidth]{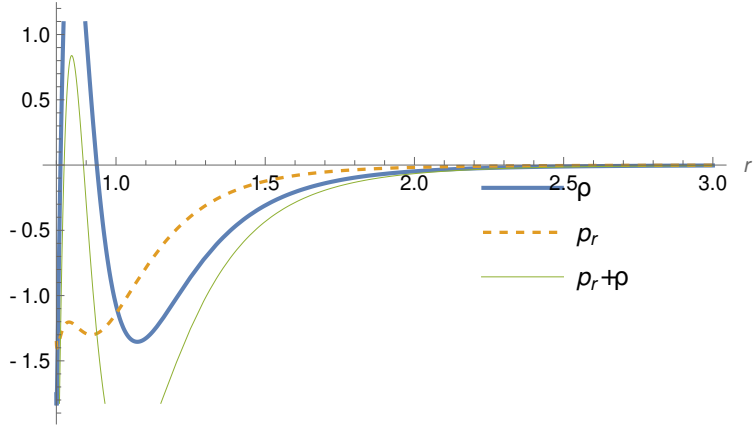}
    \caption{Plot of $\rho$, $p_r$ and $p_r+\rho$ as functions of $r$. We have set $\xi=0.05$ and $r_t=0.8$.}
    \label{fig1}
\end{figure}

Considering a throat radius such that we have ordinary matter in the throat, we plot the behavior of the energy density $\rho$, the radial pressure $p_r$, given by the Eqs.\eqref{general-energy-density} and \eqref{general-radial-pressure} respectively, and their sum $\rho+p_r$, in order to visualize the energy conditions in regions nearby the throat. The plots of $\rho$, $p_r$ and $p_r+\rho$ are presented in Fig.(\ref{fig1}).
Looking at the fig.(\ref{fig1}), we can see that we have a region nearby the throat where both energy density and the sum $p_r+\rho$ are positive, and therefore the NEC and WEC are satisfied in this region. This is a notable result due the fact that, in the general relativity limit, the NEC is never satisfied \cite{DuttaRoy:2019hij}. Therefore, in the ASG context the NEC condition is satisfied in a region nearby the throat for the case $n=4$ when we use the Ricci scalar as improvement.

\begin{figure}[ht!]
    \centering
    \includegraphics[width=0.6\textwidth]{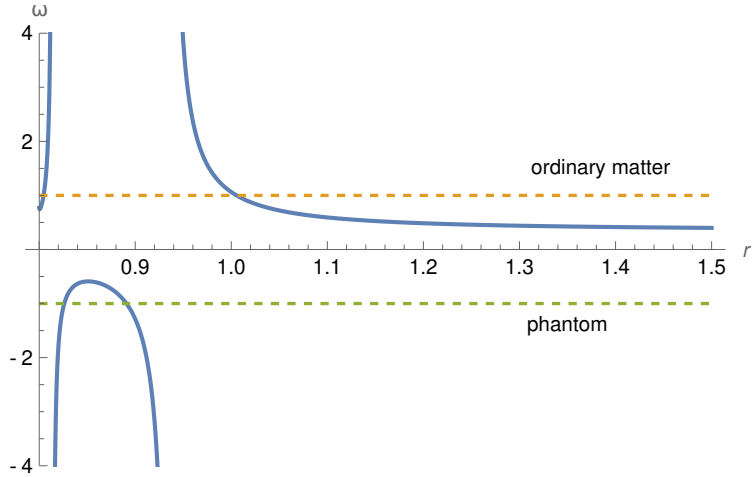}
    \caption{State parameter $\omega$ as a function of $r$. We have set $r_t=0.8$ and $\xi=0.05$.}
    \label{fig2}
\end{figure}

Also, it is instructive visualize the behaviour of the state parameter $\omega(r)$ in regions nearby the throat, in order to see what kinds of cosmological matter is necessary in other regions of space, considering again only ordinary matter in $r=r_t$. The plot of $\omega=p_r/\rho$ as a function of the radial coordinate $r$ is depicted in (fig.\ref{fig2}).

Looking at the Fig.(\ref{fig2}), we have a region of phantom-like matter ($\omega<-1$) followed, a region with quintessence-like matter ($-1<\omega<-1/3$), and then phantom again. After that, we enter in a region sourced by matter with $\omega>1$ which decays smoothly in a region of space that is sourced by ordinary matter. Once again we have a notable result, that is, the generalized solution with $n=4$ must be supported by other matters beyond Phantom (in this case, quintessence, matter with $\omega>1$ and ordinary matter) to exist. Although the necessity of extra matter beyond phantom is necessary to support these generalized geometries, as noticed by  \cite{DuttaRoy:2019hij}, they do not determine what kind of extra matter is necessary to support the wormhole geometry. Thus, the ASG brings up possible candidates of extra matter for this fact, at least for the $n=4$ case. The presence of extra matter can then be seen as a reflection of the quantum nature of matter, which in this case is introduced through the ASG.

\subsection{Squared Ricci}

Now, we use other improvement for the cutoff function, the Squared Ricci case defined by the Eq. (\ref{f2}). Firstly, we will prove the same sentence that we have when we used the Ricci scalar as improvement, that is, in the throat, we have corrections due the ASG only for the $n=4$ case. 

For this, we consider firstly the energy density given by the Eq.\eqref{IEQ-tt-s}. We will do the analysis term by term, which we have named for the sake of brevity as the terms $\rho_I$, $\rho_{II}$ and $\rho_{III}$, such that the energy density is given by $\kappa\rho=\rho_I+\rho_{II}+\rho_{III}$.

The term $\rho_I$ of the energy density is given by 
\begin{equation}
\rho_I = (1+f)\frac{b'}{r^{2}},
\end{equation}
which is easy to see that this term will give the same result for all generalized solutions with $n\geq4$. To note this, using the shape function \eqref{shape-function} and differentiating with respect to $r$ we get
\begin{equation}
b'(r)=1-(3-2n)r^{2-2n}(r^{n}-r_{t}^{n})^{2-2/n}-r^{2-n}(2n-2)(r^{n}-r_{t}^{n})^{1-2/n},
\end{equation}
which, calculating in the throat, it is easy to see that this term will give a different value only for $n=2$, providing $b'(r_t)=-1$, while gives $b'(r_t)=1$ for all $n\geq4$. The fact of the flare-out condition is not satisfied, is related to the fact that the extra matter in these generalized solutions satisfies the Averaged Null Energy Condition (ANEC) \cite{DuttaRoy:2019hij}. Writing the cutoff function $f_2=\xi\sqrt{R_{\mu\nu}R^{\mu\nu}}$ in terms of $b(r)$ we have
\begin{equation}
f_2=\frac{\xi}{2r^{3}}\sqrt{2(b+rb')^{2}+4(b-rb')^{2}},
\end{equation}
which gives a different result only for the usual case $n=2$, given by $f(r_t)=2\xi/r_t^2$, and for the generalized cases with $n\geq4$ gives the result $f(r_t)=\sqrt{2}\xi/r_t^2$, once in the throat we always have $b(r_t)=r_t$ and $b'(r_t)=1$ for all generalized solutions. Thus, the first term $\rho_I$ gives the following result in $r=r_t$ for all solutions with $n\geq4$:
\begin{equation}
\rho_I(r_t)=\frac{1}{r_t^4}(\xi\sqrt{2}+r_t^2)\,\,\,\textrm{for all}\,n\geq4.
\end{equation}

Now, we turn for the second term of the energy density $\rho_{II}$, which is given by
\begin{equation}
\rho_{II}=\left(1-\frac{b}{r}\right)\left(f_2''+\frac{2}{r}f_2'\right)=\left(1-\frac{b}{r}\right)f_2''+\left(1-\frac{b}{r}\right)\frac{2}{r}f_2'.
\end{equation}
Using analagous arguments of the Ricci scalar case, we can show that the second term of $\rho_{II}$ will contributes in the throat only for the usual Ellis-Bronnikov spacetime, i.e, for $n=2$, being zero for all $n\geq4$. Furthermore, the first term of $\rho_{II}$ gives a non null term in $r=r_t$ only for the case $n=4$, resulting in the following expression for $\rho_{II}$ in the throat
\begin{equation}
\rho_{II}(r_t)=\frac{12 \sqrt{2} \xi }{r_t^4}\,\,\,\textrm{for}\,n=4,
\end{equation}
and is zero for all $n>4$.

Finally we turn for the term $\rho_{III}$ of the energy density, which is given by
\begin{equation}
\rho_{III}=\frac{b'r-b}{2r^2}f_2'.
\end{equation}
After arranging the terms into powers of n, this term gives the following expression
\begin{eqnarray}
\rho_{III} &=&\{\xi  (24 n^4 r^{2-6 n} r_t^{3 n} (r^n-2 r_t^n) (r^n-r_t^n)^{2-\frac{6}{n}}+n^3 (8 r^{2-6 n} r_t^{2 n} (r^n-r_t^n)^{2-\frac{6}{n}} (r^{2 n}-8 r^n r_t^n+22 r_t^{2 n})\nonumber\\
&&-8 r^{-4 n} r_t^{2 n} (r^n-2 r_t^n) (r^n-r_t^n)^{\frac{n-4}{n}})+n^2 (8 r^{2-6 n} r_t^{3 n} (r^n-28 r_t^n) (r^n-r_t^n)^{2-\frac{6}{n}}-32 r^{-4 n} r_t^{3 n} (r^n-r_t^n)^{1-\frac{4}{n}})\nonumber\\
&&+16 r^{-4 n} r_t^n (2 r^n-r_t^n) (r^n-2 r_t^n) (r^n-r_t^n)^{\frac{n-4}{n}}-16 r^{-2 (n+1)} r_t^n (r^n-r_t^n)^{\frac{n-2}{n}}\nonumber\\
&&-16 r^{2-5 n} r_t^n (r^{2 n}-4 r^n r_t^n+6 r_t^{2 n}) (r^n-r_t^n)^{2-\frac{6}{n}}+n (16 r^{-2 (n+1)} r_t^n (r^n-r_t^n)^{\frac{n-2}{n}}+8 r^{2-6 n} r_t^n (-9 r^{2 n} r_t^n+2 r^{3 n}\nonumber\\
&&+16 r^n r_t^{2 n}+12 r_t^{3 n}) (r^n-r_t^n)^{2-\frac{6}{n}}-8 r^{-4 n} r_t^n (4 r^{2 n}-11 r^n r_t^n+2 r_t^{2 n}) (r^n-r_t^n)^{\frac{n-4}{n}}))\}\{2 \sqrt{2} r^{-4} [-2 r (r\nonumber\\
&&-r^{3-2 n} (r^n-r_t^n)^{2-\frac{2}{n}}) (-((2-\frac{2}{n}) n r^{2-n} (r^n-r_t^n)^{1-\frac{2}{n}})-(3-2 n) r^{2-2 n} (r^n-r_t^n)^{2-\frac{2}{n}}+1)+3 (r\nonumber\\
&&-r^{3-2 n} (r^n-r_t^n)^{2-\frac{2}{n}})^2+3 r^2 (-((2-\frac{2}{n}) n r^{2-n} (r^n-r_t^n)^{1-\frac{2}{n}})-(3-2 n) r^{2-2 n} (r^n-r_t^n)^{2-\frac{2}{n}}+1)^2]^{-1/2}\}.
\end{eqnarray}
Using the same arguments, we can see that all the terms of the type $(r^n-r_t^n)$ will not be zero in the throat only for generalized solutions with $n=4$, because for all $n>4$, all the terms $(r^n-r_t^n)$ will have a positive exponent, and therefore, vanishes in the limit $r\rightarrow r_t$. Thus, putting $n=4$ in the term $\rho_{III}$ and taking the limit $r\rightarrow r_t$ we get
\begin{equation}
\rho_{III}(r_t)=\frac{18\sqrt{2}\xi}{r_t^4}.
\end{equation}
Such term is zero for all generalized solutions with $n>4$. Finally, we have the expressions for the energy density in the throat of a generalized Ellis-Bronnikov Wormhole for the squared Ricci improvement
\begin{equation}
\kappa\rho(r_t) = \left\{
  \begin{array}{cc}
     \frac{r_t^2+7\sqrt{2}\xi}{r_t^4},  & \textrm{for}\,n=4  \\
      \frac{1}{r_t^4}(\xi\sqrt{2}+r_t^2)  & \textrm{for}\,n>4, \\  
\end{array}
\right. 
\end{equation}
where the energy density is given by $\kappa\rho=\rho_I+\rho_{II}+\rho_{III}$.

Now, we turn for the expression for the radial pressure $p_r$. Looking at the Eq.\eqref{IEQ-rr-s}, this term for a zero--tidal model is given by
\begin{equation}
\kappa p_r=-(1+f)\frac{b}{r^3}+\left(1-\frac{b}{r}\right)\frac{2}{r}f'.
\end{equation}
Note that the second term of $p_r$ is exactly the second term of $\rho_{II}$ and, as we said previously, this term is zero at the throat for all generalized solutions $n\geq4$. Therefore, only the first term of $p_r$ contributes for $n\geq4$ in $r=r_t$. Furthermore, note that, in the throat $r=r_t$, the first term of the radial pressure is exactly the negative of the term $\rho_I$, because we have $b(r_t)=r_t$ and this divided by $r_t^3$ gives $1/r_t^2$, and the cutoff function has the same expression for all $n\geq4$, as we argued previously. Therefore, we have for the radial pressure in the throat of generalized Ellis-Bronnikov wormholes when we use the squared Ricci as improvement
\begin{equation}
\kappa p_r(r_t)=-\frac{1}{r_t^4}(\xi\sqrt{2}+r_t^2) \,\,\,\textrm{for all}\,n\geq4.
\end{equation}

Once again we reached the conclusion, this time using the squared Ricci as improvement, that we have corrections at the throat due the ASG only for the case $n=4$, due the fact that the matter satisfies $p_r=-\rho$ in the throat for all $n>4$ cases. In other words, the throat for these geometries with $n>4$ must be sourced by matter with $\omega=-1$. Hence, we verify if the radial energy conditions is satisfied and the necessary cosmological matter in the throat only for the case $n=4$.

For verification of the radial energy conditions in the throat, we have for the $n=4$ case
\begin{eqnarray}
\kappa\rho(r_t)&=&\frac{r_t^2+7\sqrt{2}\xi}{r_t^4}, \\
\kappa p_r(r_t)&=&-\frac{1}{r_t^4}(\xi\sqrt{2}+r_t^2), \\
\kappa(\rho+p_r)&=&\frac{6\sqrt{2}\xi}{r_t^4}.
\end{eqnarray}
With these expressions we can see that we always have $\rho>0$ and $\rho+p_r>0$ for all values of $r_t$, and therefore the NEC and WEC conditions are satisfied in the throat, independently of the chosen values of the throat radius. Furthermore, we can easily note that we have $\rho>|p_r|$ in the throat, for all $r_t>0$, and therefore, also the DEC condition is satisfied, for all values of the throat radius. This result is quite interesting because using the squared Ricci to define the cutoff function, the radial energy conditions is always satisfied at the throat, regardless the value of the throat radius $r_t$, in contrast to what is expected by general relativity \cite{DuttaRoy:2019hij}. Also, this result is very different to what happens when we used the Ricci scalar as improvement, due the fact that the sum $\rho+p_r$ is always negative at the throat in this case, and the energy density is positive only for certain values of the throat radius $r_t$. Here, in the squared Ricci improvement, we have automatically satisfied the NEC, WEC and DEC for all $r_t>0$. With this, we can analyze what kinds of cosmological matter is necessary in the throat, calculating the state parameter $\omega=p_r/\rho$ in $r=r_t$. We have for the $n=4$ case
\begin{equation}
\omega(r_t)=\frac{r_t^2+7\sqrt{2}\xi}{-r_t^2-\sqrt{2}\xi}.
\end{equation}
Analyzing this as a function of $r_t$, we can easily see that this time we always have phantom energy in the throat ($\omega(r_t)<-1$), independently of the values of $r_t$ in terms of $\xi$, and this is a quite different result in comparison to those found for the Ricci scalar case. Interestingly, the squared Ricci model provides matter satisfying the radial energy conditions in the throat, however, the exotic matter (phantom) is always necessary in the throat, while in the Ricci scalar model the energy conditions can not be satisfied, although there is the possibility of having non-exotic matter as source in the throat, when $n=4$. Note that the squared Ricci case leads to the necessity of phantom in the throat of these generalized solutions. Therefore, this case does not cure the problem of exotic matter in the wormhole throat of generalized Ellis-Bronnikov solutions.

\begin{figure}[ht!]
    \centering
  \includegraphics[width=0.6\textwidth]{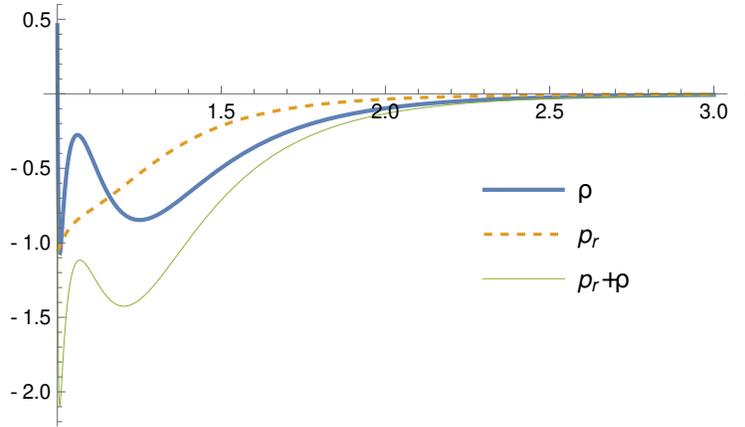}
    \caption{Energy density $\rho$, radial pressure $p_r$ and the sum $\rho+p_r$ as a function of the radial coordinate $r$ in the squared Ricci case. We have set $r_t=1$ and $\xi=0.05$.}
    \label{fig3}
\end{figure}

Although the squared Ricci case in ASG approach phantom energy is always necessary in the throat, we plotted the behavior of the energy density $\rho$, the radial pressure $p_r$, and their sum $\rho+p_r$ in the Fig.(\ref{fig3}), for $n=4$ case, in order to see if the energy conditions are satisfied in regions nearby the throat. 
Looking at the Fig.(\ref{fig3}), we can see that the radial energy conditions are not satisfied in regions nearby the throat, as well as we have in the general relativity limit \cite{DuttaRoy:2019hij}. As we can see, the regions near of the  wormhole throat with $n=4$, we have $\rho<0$ and $\rho+p_r<0$. These results, again, are in totally contrast to the results that we found when we used the Ricci scalar model.

\begin{figure}[ht!]
    \centering
    \includegraphics[width=0.6\textwidth]{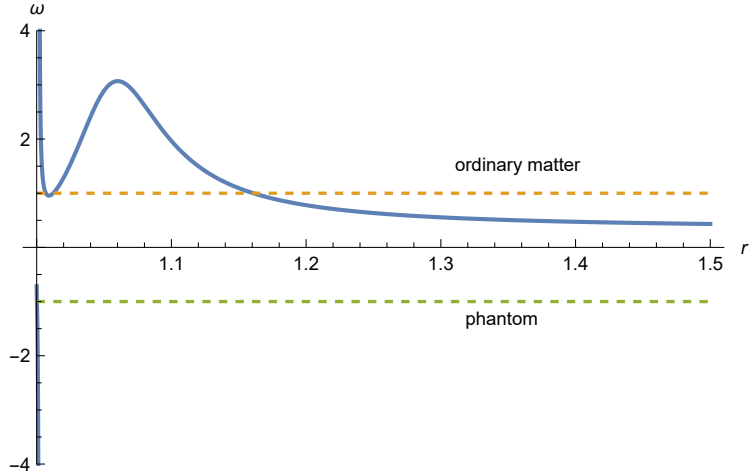}
    \caption{Parameter state $\omega$ as a function of $r$ in the squared Ricci case. We have set $r_t=1$ and $\xi=0.05$.}
    \label{fig4}
\end{figure}

Similarly to what we did for the previous case of the Ricci scalar, we plot the behavior of the state parameter $\omega(r)$ for the case $n=4$ in regions near of the throat, using the squared Ricci as improvement. Remember that, in this case, we always have phantom energy in the throat, whatever the value that we choose for the throat radius $r_t$. Interestingly, looking at the graph of $\omega(r)$ depicted in Fig.(\ref{fig4}), in this case the necessity of phantom energy is only in the throat, while in regions near the throat is necessary matter with $\omega(r)>1$ and ordinary matter. Once again, the ASG brings naturally other kinds of matter beyond phantom as source for the generalized Ellis-Bronnikov wormhole, at least for the $n=4$ case. The necessity for these ``extra matter'' to support this family of wormholes was noticed by \cite{DuttaRoy:2019hij}, and one more time, the ASG brings the possible candidates for these extra matter, at least for the $n=4$ case. Again, we can think that the necessity for these ``extra matter'' is due the quantum nature of the matter.

\subsection{Kretschmann scalar}

This time we analyze the improvement made by the Kretschmann scalar, which for the generalized Ellis-Bronnikov wormholes is given by the Eq.\eqref{f3}. Just like we did before, let us analyze the energy density and radial pressure expressions in the throat using this cutoff function.

First, let us talk about the energy density. As we did in the case of the squared Ricci model, we will analyze the energy density term by term due in order to facilitate the analysis. Again, we will divide the energy density in three terms that we call $\rho_I$, $\rho_{II}$ and $\rho_{III}$, respectively, such that $\kappa\rho=\rho_I+\rho_{II}+\rho_{III}$.

The term $\rho_{I}$ of the energy density is given by the expression
\begin{equation}
\rho_{I}=(1+f_3)\frac{b'}{r^2},
\end{equation}
and, as we said, we always have $b'(r_t)=1$ for all generalized wormholes, that is, for all the cases $n\geq4$, and, writing the cutoff function $f_3$ as a functional of the $b(r)$, $f_3$ it will be written in terms of $b(r)$ and their derivatives, and therefore, in the throat, it gives the same result for all generalized solutions. In the case of the Kretchsmann scalar, $f_3$ is written in terms of $b(r)$ by the expression
\begin{equation}
f_3=\xi\sqrt{\frac{2 r^2 b'(r)^2-4 r b(r) b'(r)+6 b(r)^2}{r^6}},
\end{equation}
and calculating in $r=r_t$ and using the facts $b(r_t)=r_t$ and $b'(r_t)=1$, which is valid for all $n\geq4$ cases, we found
\begin{equation}
f_3(r_t)=\frac{2\xi}{r_t^2}\,\,\,\textrm{for all}\,n\geq4, 
\end{equation}
and this results in the following expression for the term $\rho_I$ of the energy density in $r=r_t$
\begin{equation}
\rho_{I}(r_t)=\left(1+\frac{2\xi}{r_t^2}\right)\frac{1}{r_t^2}=\frac{1}{r_t^4}(r_t^2+2\xi), \,\,\,\textrm{for all}\,n\geq4.
\end{equation}

For the term $\rho_{II}$ of the energy density, we have
\begin{equation}
\rho_{II}=\left(1-\frac{b}{r}\right)\left(f_3''+\frac{2}{r}f_3'\right).
\end{equation}
We follow the same line of reasoning when we used the squared Ricci model in order to facilitate the analysis. The result is that, as well as in the squared Ricci model, the term with the first derivative of the cutoff function will have contribution in the throat only for the usual Ellis-Bronnikov case $n=2$, and will be zero for all generalized cases $n\geq4$. However, unlike the result obtained in the squared Ricci model, the term with second derivative of the cutoff function also contributes in the throat only for the usual Ellis-Bronnikov wormhole. Therefore, for all generalized Ellis-Bronnikov wormholes, we have

\begin{equation}
\rho_{II}(r_t) = 0 \,\,\textrm{for all}\,n\geq4.
\end{equation}

Finally, we analyze the expression for the term $\rho_{III}$ of the energy density which is given by
\begin{equation}
\rho_{III}=\frac{b'r-b}{2r^2}f_3', 
\end{equation}
which using the cutoff function $f_3$ we have the following equation for the term $\rho_{III}$
\begin{eqnarray}
\rho_{III} &=&4 (n-1) \xi  r^{-6 n-4} r_t^n (r^n-r_t^n)^{1-4/n} \{(r^{2 n+2} r_t^{2 n} ((n^3+5) r^2-(n+1) (r^n-r_t^n)^{2/n})-(n (n (3 n-4)+2)+3) r^{n+4} r_t^{3 n}\nonumber\\
&&+n (2 (n-2) n+3) r^4 r_t^{4 n}+(n+3) r^{3 n+2} r_t^n(r^n-r_t^n)^{2/n}-r^2)+r^{4 n} (r^2-(r^n-r_t^n)^{2/n})^2)\}\nonumber\\
&&\{[r^{-4 (n+1)} (-4 ((n-2) n+2) r^{n+4} r_t^{3 n}+(2 (n-2) n+3) r^4 r_t^{4 n}+4 r^{3 n+2} r_t^n ((r^n-r_t^n)^{2/n}-r^2)\nonumber\\
&&+r^{4 n} (r^2-(r^n-r_t^n)^{2/n})^2+2 r^{2 n+2} r_t^{2 n} (((n-2) n+4) r^2-(r^n-r_t^n)^{2/n}))]^{-1/2}\}.
\end{eqnarray}
 
Again, this term is zero in the throat for all generalized Ellis-Bronnikov wormholes, that is, for all $n\geq4$ cases. This is due the fact of, for $n>4$, the term $(r^n-r_t^n)^{1-4/n}$ multiplying all the terms between the braces will have a positive exponent, and, in the limit $r\rightarrow r_t$, this term is zero, vanishing all the terms in braces, yielding a zero result for the term $\rho_{III}(r_t)$. And, if $n=4$, although this brings a zero for the exponent, $(r^n-r_t^n)^0$, giving the unity, the terms between braces will be zero, as it can be easily checked. Therefore, the term $\rho_{III}$ of the energy density is also zero in $r=r_t$ for all generalized Ellis-Bronnikov wormholes when we use the Kretschmann scalar as improvement. Then, we have
\begin{equation}
\rho_{III}(r_t)=0\,\,\textrm{for all}\,n\geq4.
\end{equation}
As the energy density is given by $\kappa\rho=\rho_I+\rho_{II}+\rho_{III}$, only the term $\rho_I$ contributes in the throat for $n\geq4$ cases, giving the result
\begin{equation}
\kappa\rho(r_t)=\frac{1}{r_t^4}(r_t^2+2\xi)\,\,\,\textrm{for all}\,n\geq4.
\end{equation}
The Kretchsmann scalar is the only invariant considered here that gives the same result for the energy density in $r=r_t$ for all the generalized Ellis-Bronnikov wormholes.

Now, using the expression for the radial pressure, we have
\begin{equation}
\kappa p_r=-(1+f_3)\frac{b}{r^3}+\left(1-\frac{b}{r}\right)\frac{2}{r}f_3'.
\end{equation}
We can note that, similar to the previous case, the second term of the radial pressure is equal to the term of the first derivative of $f$ of the term $\rho_{II}$, and, as we said, this terms is always zero for $n\geq4$ in the throat when we use the Kretschamnn scalar. The first term of the radial pressure in the throat gives exactly the negative of the term $\rho_I$ of the energy density, because $b(r_t)=r_t$, and dividing this for $r_t^3$ gives a multiplicative factor $1/r_t^2$, and the $f_3(r_t)$ we show that this gives $f_3(r_t)=2\xi/r_t^2$ for all generalized cases. Therefore, the radial pressure have the same result for all $n\geq4$ and we have
\begin{equation}
\kappa p_r=-\frac{1}{r_t^4}(2\xi+r_t^2) \,\,\,\textrm{for all}\,n\geq4,
\end{equation}
Interestingly, when we use the Kretschmann scalar to define the cutoff function, the necessary matter in the throat satisfies the state equation $p_r(r_t)=-\rho(r_t)$ for all generalized cases, that is, in $r=r_t$ we always have a matter with $\omega(r_t)=-1$ in the throat. This is the same result reported in \cite{Kar:1995jz,DuttaRoy:2019hij}.

\section{Final remarks}

In this paper we study a class of wormhole solutions called generalized Ellis-Bronnikov wormholes in the context of asymptotically safe gravity. These solutions are characterized by two parameters: an even number $n$ and the wormhole throat radius $r_t$. The particular case $n=2$ recovers the usual Ellis-Bronnikov spacetime, which has already been addressed in the literature. We analyzed the nature of matter in the wormhole's throat, and in nearby regions, of these generalized solutions with $n>2$, using three curvature scalars in the ASG approach, namely, the Ricci scalar, squared Ricci and the Kretschmann scalar. We have shown that the ASG leads to corrections in the matter at the wormhole's throat only for the $n=4$ case.

Using the Ricci scalar to define the cutoff function, we show that the ASG approach leads to corrections in the throat only for the generalized case with $n=4$, where we have the possibility that the throat can be sourced by various types of matter, including ordinary matter. This is contrary to what is predicted by general relativity, where necessarily we have phantom as source. Also, in this case, there is the possibility of having matter satisfying the radial energy conditions, as we saw in Fig.(\ref{fig1}). This feature is remarkable, since in the general relativity the NEC is never satisfied \cite{DuttaRoy:2019hij}. In addition, looking at the Fig.(\ref{fig2}), it is necessary various types of matter beyond phantom when we consider other regions of space. Thus, the ASG method brings naturally possible candidates of extra matter that these generalized solutions must have in order to support their existence, as pointed in \cite{DuttaRoy:2019hij}.

However, when we use the squared Ricci model, the results are quite different from what we expect. Although in this case we have corrections in the throat only for the case $n=4$ too, the presence of phantom energy is always necessary in the throat, and therefore, the problem of the exotic matter in the throat can not be solved. However, in this case we have the possibility of having all radial energy conditions being satisfied at the throat. Unlike the Ricci scalar, the energy conditions are not satisfied in other regions of space, as predicted by general relativity \cite{DuttaRoy:2019hij}. Furthermore, when we analyze the state parameter as the function of $r$, although phantom is necessary in the throat, once again we have the necessity of extra matter beyond phantom as source of the wormhole spacetime, including ordinary matter.

Using the Kretschmann scalar as improvement, we have no corrections due the ASG, that is, the throat necessarily must be sourced by matter that satisfies the state equation $p_r=-\rho$, i.e, we always have matter with $\omega=-1$, as well as reported in \cite{Kar:1995jz,DuttaRoy:2019hij}.

Then, we see that for the curvature invariants considered here, only the Ricci scalar provides the possibility of having only ordinary matter in the throat, and only for the case with $n=4$. When we use the squared Ricci, we have corrections only for the case $n=4$ too, but the presence of phantom in the throat is always necessary. The Kretschmann scalar reproduce the same results reported in \cite{Kar:1995jz,DuttaRoy:2019hij}. Therefore, in the ASG context, the throat without phantom is observed only for the Ricci scalar and the Kretschmann scalar cases. The Ricci scalar provides several possibilities of matter in the throat beyond phantom, including ordinary matter. Then, the generalized Ellis-Bronnikov wormholes yields to the possibility of having only ordinary matter at the throat in the context of asymptotically safe gravity.

\section*{Acknowledgments}

The authors would like to thank Conselho Nacional de Desenvolvimento e Tecnológico (CNPQ), Coordenação de Aperfeiçoamento de Pessoa de Nível Superior - Brasil (CAPES) and Fundação Cearense de Apoio ao Desenvolvimento Científico e Tecnológico (FUNCAP).

\end{document}